\documentclass[preprint,12pt]{elsarticle}
\usepackage[utf8]{inputenc}
\usepackage{amssymb}
\usepackage{amsmath}
\usepackage{makecell}
\usepackage{xcolor}
\usepackage{graphicx}
\newcommand{\etal}{\textit{et al}. }
\usepackage[a4paper, left=2cm, right=2cm, top=2.5cm, bottom=2.5cm]{geometry}
\usepackage{microtype}
\usepackage{tabularx}
\usepackage[numbers]{natbib}
\definecolor{mypurple}{RGB}{114, 9, 183}
\journal{Nuclear Physics B}
\begin{document}
\begin{frontmatter}

\title{Integrating Biological and Machine Intelligence: Attention Mechanisms in Brain-Computer Interfaces}

\author{Jiyuan Wang\textsuperscript{a,b,1},
Weishan Ye\textsuperscript{a,b,1}, Jialin He\textsuperscript{a,b}, Li Zhang\textsuperscript{a,b}, Gan Huang\textsuperscript{a,b},\\ Zhuliang Yu\textsuperscript{c,d}, and Zhen Liang\textsuperscript{a,b,*}} %% Author name

% Author affiliation
\affiliation{organization={The School of Biomedical Engineering},
            addressline={Medical School}, 
            city={Shenzhen University},
            state={Shenzhen},
            country={China}}
\affiliation{organization={The Guangdong Provincial Key Laboratory of Biomedical Measurements and Ultrasound Imaging},
            state={Shenzhen},
            country={China}}
\affiliation{organization={Shien-Ming Wu School of Intelligent Engineering},
            addressline={South China University of Technology}, 
            state={Guangdong},
            country={China}}
\affiliation{organization={Institute for Super Robotics},
            state={Guangdong},
            country={China}}

% Abstract
\begin{abstract}
With the rapid advancement of deep learning, attention mechanisms have become indispensable in electroencephalography (EEG) signal analysis, significantly enhancing Brain-Computer Interface (BCI) applications. This paper presents a comprehensive review of traditional and Transformer-based attention mechanisms, their embedding strategies, and their applications in EEG-based BCI, with a particular emphasis on multimodal data fusion. By capturing EEG variations across time, frequency, and spatial channels, attention mechanisms improve feature extraction, representation learning, and model robustness. These methods can be broadly categorized into traditional attention mechanisms, which typically integrate with convolutional and recurrent networks, and Transformer-based multi-head self-attention, which excels in capturing long-range dependencies. Beyond single-modality analysis, attention mechanisms also enhance multimodal EEG applications, facilitating effective fusion between EEG and other physiological or sensory data. Finally, we discuss existing challenges and emerging trends in attention-based EEG modeling, highlighting future directions for advancing BCI technology. This review aims to provide valuable insights for researchers seeking to leverage attention mechanisms for improved EEG interpretation and application.
\end{abstract}


\end{frontmatter}

\footnotetext[1]{\hspace{1mm}Equal contributions.}

\section{Introduction}
\label{introduction}
%% Labels are used to cross-reference an item using \ref command.

Research in brain-computer interfaces (BCIs) has long been challenged by the need to process large, complex datasets of brain signals \cite{chen2017brain}. The primary difficulty arises from the diverse and complicated nature of electroencephalography (EEG) signals, which requires efficient and effective strategies for signal analysis and modeling \cite{subha2010eeg}. Attention mechanism-based models have shown exceptional promise in tackling the complexities of EEG signal processing \cite{abibullaev2023deep}. By selectively focusing on critical information within extensive brain signal datasets, Attention models help to minimize irrelevant noise, thereby significantly improving data processing efficiency \cite{de2022attention}. Attention mechanisms not only enhances the effectiveness of BCI research but also introduces greater flexibility and intelligence in the development of models tailored for BCI applications \cite{keutayeva2023exploring}.

Attention mechanisms draw inspiration from biological visual and auditory processes, as well as cognitive processes in psychology \cite{fukui2019attention}. Research has demonstrated that during visual and auditory recognition tasks, humans naturally focus on key elements while suppressing irrelevant information, thereby enhancing the accuracy and efficiency of recognition and decision-making \cite{soydaner2022attention}. Leveraging this principle, attention mechanisms in models, commonly referred to as attention models, are designed to assign flexible weights to different features \cite{lv2022attention}. This enables the model to concentrate on critical information pertinent to the target task while filtering out extraneous data \cite{niu2021review}. Attention models also enhance the understanding of the relationships between input and output data, which in turn improves the model's interpretability \cite{gao2021interpretable}. This enhancement not only maximizes the effective utilization of data but also reduces the impact of data variability caused by individual differences, as the model can learn to prioritize more representative features. In BCI research, these capabilities are particularly valuable, as they help enhance the accuracy of neural decoding, improve the robustness of brain modeling, and enable more adaptive and personalized brain-computer interaction \cite{wang2023mtrt}. Furthermore, attention models are well-suited for multimodal BCI applications, where they facilitate the efficient fusion of features extracted from different modalities \cite{DUAN2024102536}. As a result, the use of attention models in BCI-related research tasks holds significant potential and offers promising avenues for further exploration.

\begin{figure*}
\begin{center}
\includegraphics[width=1\textwidth]{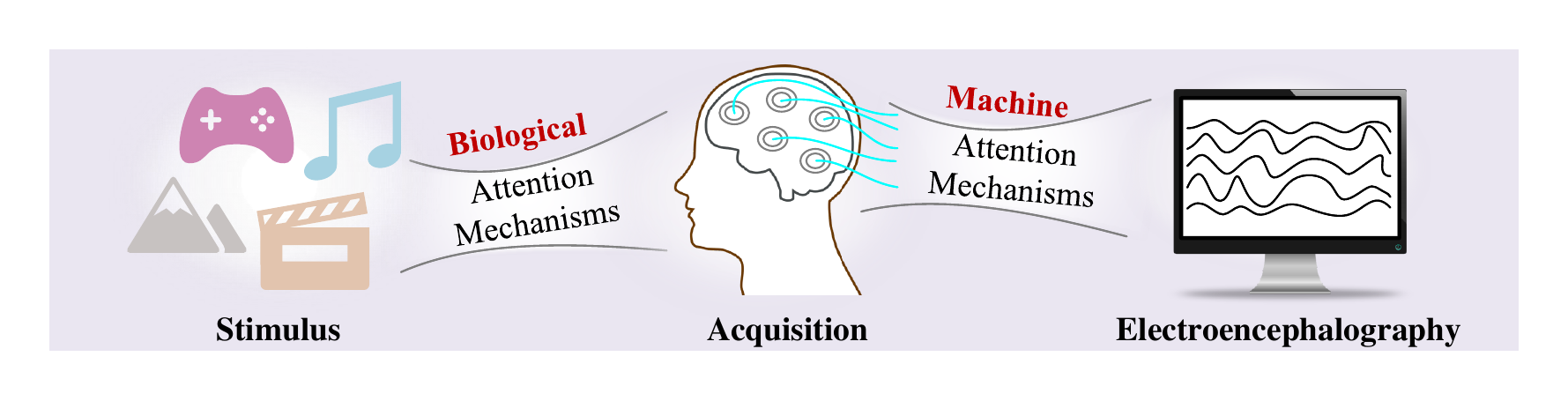}
\end{center}
\caption{The interaction between BCI and attention mechanisms}
\label{fig:global_picture}
\end{figure*}

Attention models have been initially applied extensively in computer vision and natural language processing (NLP) domains, typically integrated as modules within the backbone frameworks of convolutional neural networks (CNNs) and recurrent neural networks (RNNs) \cite{xie2023attention}. In 2014, Mnih \etal \cite{mnih2014recurrent} and Bahdanau \etal \cite{bahdanau2014neural} introduced attention mechanisms in RNNs for image classification and machine translation tasks in NLP, respectively. The introduction of a novel self-attention mechanism by Vaswani \etal \cite{vaswani2017attention} in 2017, through the "Transformer" model architecture for machine translation tasks, further propelled the use of attention models. Since then, Transformer models and their variants have been applied across various tasks \cite{islam2023comprehensive}. For example, Dosovitskiy \etal proposed the Vision Transformer \cite{dosovitskiy2020image}, demonstrating that a pure Transformer architecture could be effective for computer vision tasks without relying on CNN modules. Additionally, Liu \etal introduced the Swin Transformer \cite{liu2021swin}, which utilizes a windowed self-attention mechanism to reduce the computational complexity of Transformer models. Nowadays, attention mechanisms are foundational to modern deep learning, with their flexibility and efficacy continuing to revolutionize artificial intelligence research and applications.

\begin{figure*}
\begin{center}
\includegraphics[width=1\textwidth]{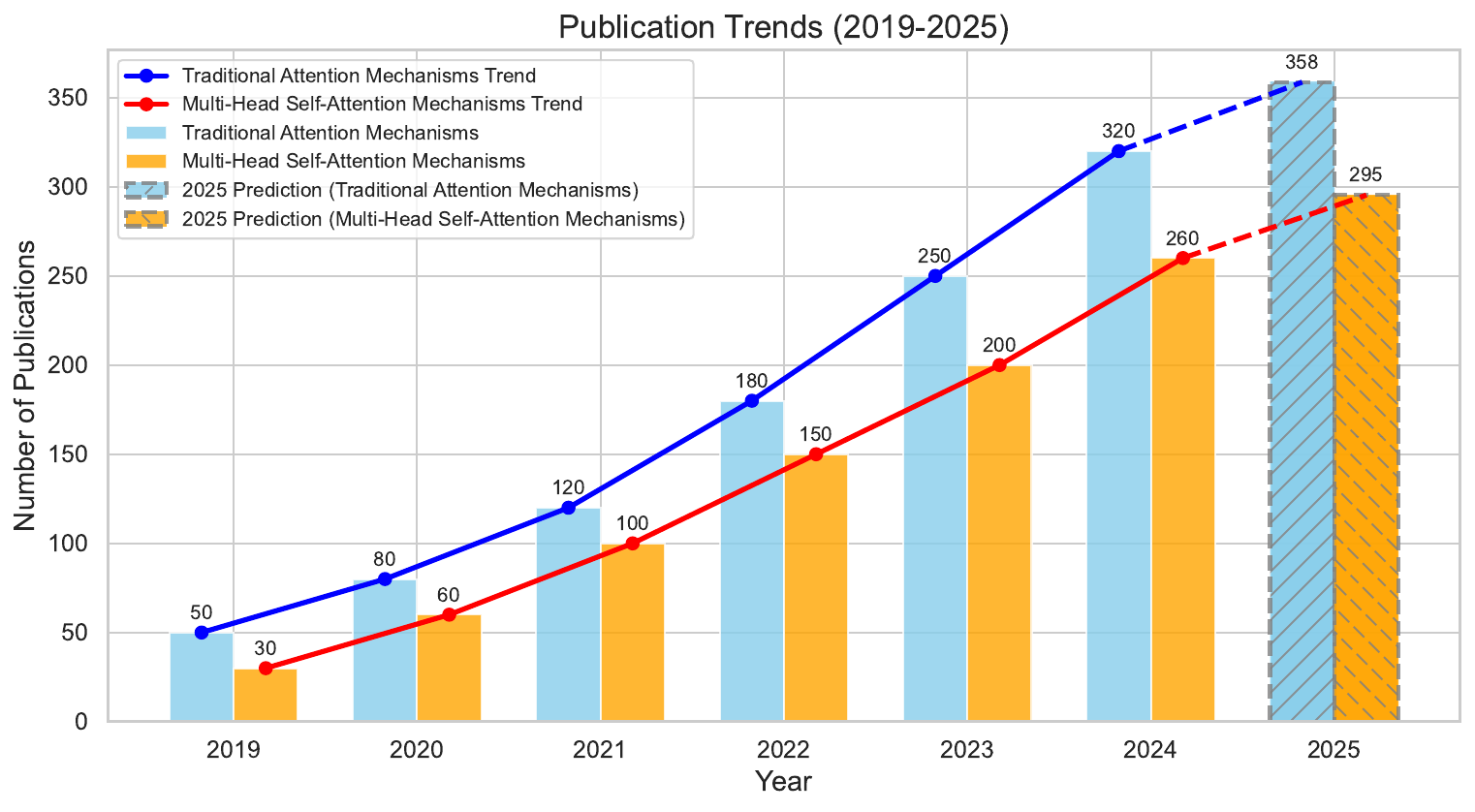}
\end{center}
\caption{The number of papers retrieved from Google Scholar for the two keyword combinations.}
\label{fig:papers}
\end{figure*}

Building on significant breakthroughs in computer vision and NLP, attention models have also attracted substantial interest in the BCI field, catalyzing rapid progress in integrating EEG signal processing with attention mechanisms. To assess the growing interest in this area, we conduct a literature search using Google Scholar to track the number of publications since 2019. The search is performed using two keyword combinations: (1) "attention"+"EEG"+"deep learning", and (2) "attention"+"EEG"+"Transformer"
+"deep learning".The searching results are presented in Fig. \ref{fig:papers}, showing the number of papers retrieved from Google Scholar for each of the two keyword combinations. In the BCI domain, attention mechanism modeling generally falls into two categories.  \textbf{(1) Traditional Attention Mechanism-Based Modeling.} It calculates attention weights for various types of information in EEG signals, such as spatial, temporal, and spectral features, prioritizing those most relevant to the task. \textbf{(2) Transformer-Based Multi-Head Self-Attention Modeling.} It employs multiple attention heads to simultaneously focus on different parts of the EEG data, enabling the model to capture both global and local relationships across various dimensions \cite{chen2022exploring}. Furthermore, extending these two modeling strategies to multimodal applications significantly enhances the model's ability to process and integrate information from different modalities \cite{vortmann2022multimodal}. Such integration is particularly critical for developing efficient and accurate BCI systems, as it allows for a more comprehensive understanding of the user's intentions and mental state.

The following sections provide a comprehensive exploration of the applications of attention models in BCIs, focusing on their role in enhancing the understanding of EEG signals and advancing BCI technology. Section 2 introduces the concept of traditional attention mechanisms and categorizes their specific applications in EEG signal modeling. Section 3 details EEG signal modeling methods based on Transformer multi-head self-attention mechanisms. Given the growing popularity of multimodal models, Section 4 discusses the application of attention models in multimodal contexts. Finally, Section 5 summarizes the key points of this work and provides future perspectives on the use of attention mechanisms in EEG signal modeling.

\section{Traditional Attention Mechanisms in EEG}
%% Use \subsection commands to start a subsection.
Traditional attention mechanism-based modeling enhances performance and generalization by efficiently selecting features through adaptive weighting and combining different types of information. Given input data, attention modeling dynamically computes feature weights based on prior knowledge or task-specific requirements.  Depending on how these weights are applied, attention mechanisms can be broadly categorized into soft and hard attention. In the soft attention mechanism, each feature is assigned a weight that is continuously distributed between 0 and 1. These weights are differentiable, which allows them to be optimized through continuous learning within the network model \cite{shen2018reinforced} Compared to hard attention, where features are either entirely selected or ignored, soft attention provides a more refined weighting approach, enabling the model to learn the relative importance of features more effectively. This leads to smoother gradient flow during backpropagation, contributing to more stable and efficient training \cite{lu2023multi}. In contrast, hard attention assigns non-differentiable weights, which cannot be optimized through conventional deep learning techniques \cite{chen2021deep} Due to these limitations, hard attention is challenging to integrate directly with traditional deep learning models. Therefore, this paper will not cover or summarize research work related to hard attention mechanisms.

Building on this foundational understanding of attention mechanisms, it is crucial to explore their implementation in practical scenarios, particularly with EEG data. Attention modules can vary significantly depending on their scope and integration into the model, and analyzing these variations provides a more comprehensive understanding of their impact. To do so, we will focus on two critical aspects of attention module implementation: the specific types of attention modules used and their methods of embedding within broader model architectures.

\subsection{Types of Attention Modules}
\begin{figure*}[h]
\centering
\includegraphics[width=1\textwidth]{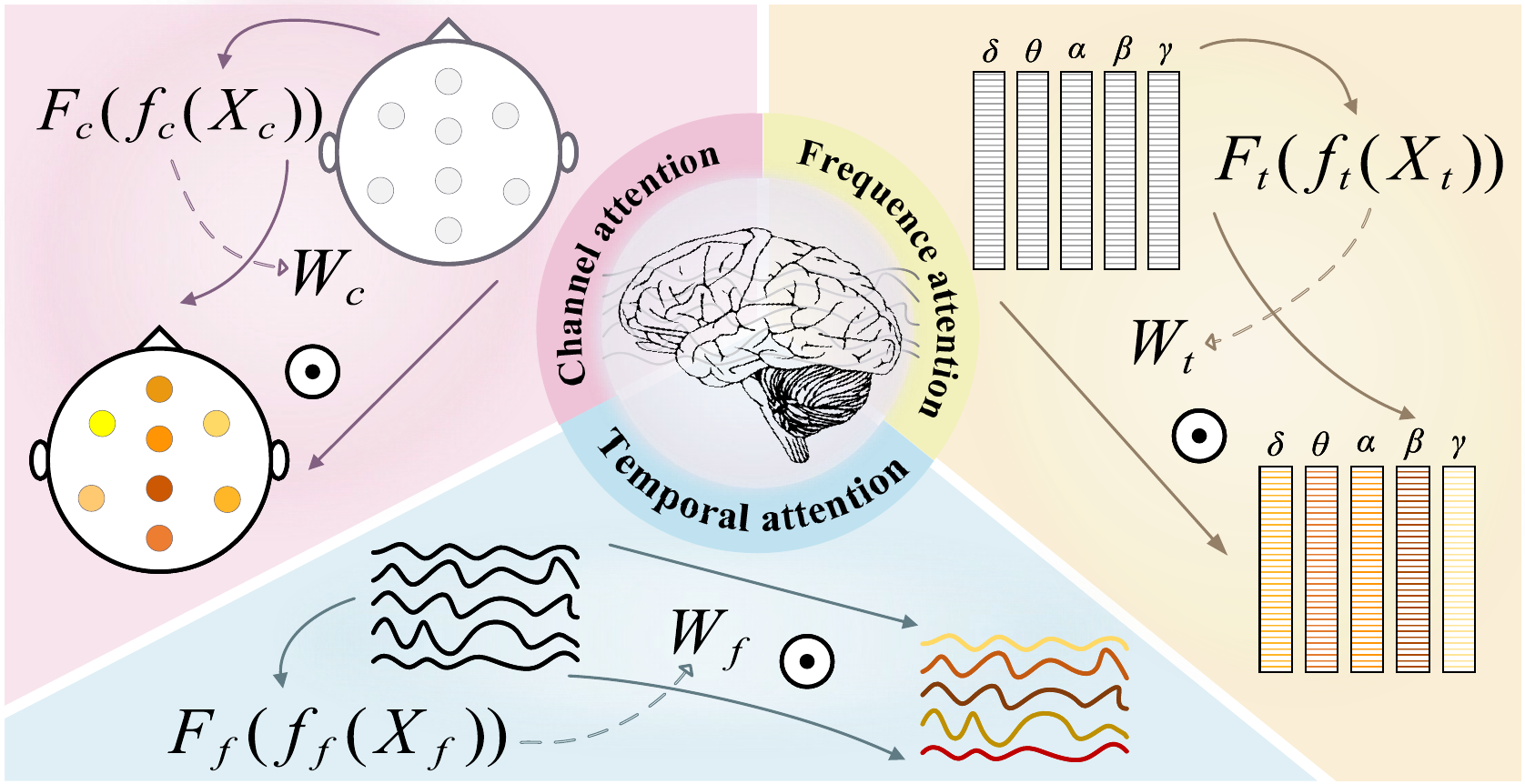}
\caption{Types of attention modules in traditional attention mechanism-based modeling. The attention weights are defined as $W_c = F_c(f_c(X_c))$ for channel attention, $W_t = F_t(f_t(X_t))$ for temporal attention, and $W_f = F_f(f_f(X_f))$ for frequency attention. Here, $f(\cdot)$ represents a transformation applied by the model to the input, and $F_n(\cdot)$ denotes a normalization function, commonly implemented as the softmax function in practical applications.}
\label{fig:TypeAtt}
\end{figure*}

In brain modeling tasks, attention mechanisms enhance feature extraction from EEG signals across channel, temporal, and frequency dimensions by assigning weights to highlight the most relevant information, as shown in Fig. \ref{fig:TypeAtt}.

\textbf{(1) Channel Attention Module.}

The channel attention module is designed to assess and adaptively weight the importance of individual EEG channel. The brain, as a complex biological system, relies on specialized brain regions that fulfill distinct roles while maintaining inter-regional interactions to adapt to diverse real world tasks\cite{toulmin2015specialization}. Extensive research from neuroscience and computational studies indicates that different brain areas contribute unequally to specific cognitive functions. In BCI applications, channel attention module effectively identifies and emphasizes the most informative brain regions for tasks such as motor imagery, emotion recognition, and visual perception\cite{wu2023classification, chen2021multiattention, tao2020eeg, liu20213dcann,Li2023MESNP,nagarajan2023relevance,hild2010optimal}. By assigning weights to EEG channels based on their task relevance, it allows models to focus on salient signals from informative channels and suppress noises from less informative channels. The most straightforward approach to implement the traditional attention mechanism to the channel dimension of electroencephalography (EEG) signals is to initialize trainable weight parameters for each channel. These weights are multiplied with the corresponding EEG input features and dynamically updated during model training, Higher weights signify greater channel importance for the task. For example, Huang \etal introduced a channel attention layer within a CNN-Bi-LSTM architecture to balance contributions of EEG channels in emotion recognition\cite{huang2023model}. Su \etal designed a soft attention mechanism that generates channel masks by optimizing task objectives for auditory attention detection in multi-speaker scenarios\cite{su2021auditory}. Hsu \etal proposed the Deep EEG-Channel-Attention (DEC) module, which adaptively adjusts channel weights in motor imagery classification\cite{hsu2023eeg}. Similar efforts by Wimpff \etal and Zhang \etal further validate the utility of channel attention mechanism\cite{wimpff2023eeg,zhang2021affective}.
 Mathematically, given a multi-channel EEG input $ X_c \in \mathbb{R}^{ C \times T}$($C$ denotes the number of channels and $T$ is the number of time points), a trainable attention weight vector $W_c \in \mathbb{R}^{1 \times C}$ is introduced. The weighted output $\hat{X}_c$ is computed as:
\begin{equation}
\hat{X}_c= \sigma(X_c \odot W_c) 
\end{equation}
where $\sigma$ denotes an activation function (e.g., Sigmoid or Softmax), and $\odot$ denotes element-wise multiplication.

To capture the spatial dependencies between EEG electrodes, channel attention can be integrated with Graph Neural Networks (GNNs) \cite{scarselli2009graph}. GNNs are particularly well-suited for EEG data due to their ability to model non-Euclidean structures, such as the irregular spatial layout of scalp electrodes. They represent EEG channels as nodes and their spatial relationships as edges in a graph. For example, Lin \etal demonstrated that retaining only 20\% of EEG channels via attention-based selection within a GNN architecture yielded over 90\% accuracy on the DEAP and SEED datasets. Even when limited to the top 12 channels, performance declined by only 1.3\% on SEED \cite{lin2023eeg}. Additional channel-specific weighting modules can be introduced prior to GNN input, further enhancing flexibility and performance\cite{chen2025graph}. The general formulation for channel attention in GNNs is as follows:
\begin{equation}
H(x) = \sigma(\tilde{\mathbf{A}}F(X\odot W_c)W_G)
\end{equation}
Here, $\sigma$ represents an activation function (e.g. sigmoid or softmax), $\tilde{\mathbf{A}}$ is the adjacency matrix, $ F(.)$ is a mapping function (e.g., convolutional or fully connected layer), and $W_c$ is a learnable GNN parameter, and the rest are as previously defined.

Recent advances in channel-specific attention mechanisms for EEG have focused on enhancing spatial modeling without relying on temporal or spectral features. By concentrating on channels associated with key brain functions, these mechanisms improve feature extraction and boost the performance of brain-computer interface systems and other neural decoding applications. A notable innovation is the use of dynamic, data-driven channel connectivity, where traditional fixed adjacency matrices based on anatomical electrode positions are replaced by adaptive graph structures learned directly from data. This flexibility allows the model to assign edge weights via attention scores that reflect task-specific functional relevance. For example, it emphasizes frontal-parietal connections during emotional processing, which could capture context-dependent neural couplings unique to each individual or task. Another promising direction involves sparse and interpretable channel weighting through regularization techniques such as L1-norm, which encourage the selection of a minimal subset of critical channels. This reduces redundancy while aligning with established neurophysiological insights, enhancing model transparency. Hierarchical channel grouping provides further refinement by organizing electrodes into brain regions and applying attention at both inter-regional and intra-regional levels, effectively capturing multi-scale spatial dependencies in neural activity. To address inter-subject variability, cross-subject adaptive attention mechanisms incorporate meta-learning or transfer learning strategies to personalize channel weighting for new individuals, leveraging common neural topologies while accommodating unique signal characteristics. Finally, lightweight attention architectures have been developed to ensure computational efficiency in real-time applications by incorporating compact modules such as bottleneck layers that select essential channels with minimal latency. Collectively, these innovations significantly advance the modeling of EEG spatial dynamics, enabling more adaptive, interpretable, and efficient neural systems.

\textbf{(2) Temporal Attention Module.} 

The temporal attention module is designed to effectively capture the dynamic fluctuations in brain activity over time, acknowledging that EEG signals reflect temporally varying neural states depending on the nature of the task. During task execution, cognitive processes unfold across multiple phases, with certain moments carrying higher relevance while others contribute less informative or even distracting signals. For example, in emotion recognition tasks, the onset, peak, and offset of emotional responses are distributed across distinct time points, reflecting task-specific temporal patterns\cite{zhu2015time,thiruchselvam2011temporal}. Findings in neuroscience and cognitive science have consistently demonstrated that different time steps in EEG signals contribute unevenly to cognitive function processing\cite{wairagkar2021temporal,angrilli2003temporal,yu2024erp}. By assigning higher attention weights to task-relevant time intervals and suppressing irrelevant fluctuations, the temporal attention module enhances the model’s capacity to extract discriminative temporal features. It ultimately improves the robustness and precision of brain activity analysis in time-series modeling.

Due to the inherently non-stationary nature of EEG signals, two fundamental challenges arise when applying temporal attention: determining the appropriate duration of each time window and defining the optimal number of segments necessary to isolate meaningful neural responses. These parameters often vary across tasks and recording conditions, making them more complex to optimize than the fixed and predefined spatial structure seen in channel attention mechanisms. To address these issues, researchers have proposed various solutions. Jiang \etal introduced a dynamic weighting strategy for identifying key temporal fragments, demonstrating that task performance, such as emotion recognition, can be significantly influenced by the selected window length\cite{jiang2022electroencephalogram}. Their work suggested that optimal segment durations should be adapted based on task characteristics, such as distinguishing between structured emotional induction and natural conversation. Yang \etal developed a Gated Temporal-Separable Attention Network that leverages causal and dilated convolutions to extract multi-scale temporal features, subsequently assigning attention weights to relevant segments \cite{yang2023gated}. Beyond these examples, temporal attention has been widely applied in diverse BCI contexts, including cognitive and emotional state monitoring\cite{zhang2022eeg, jia2020sst} as well as motor imagery classification \cite{zhang2020motor, ma2022time}. A general representation of temporal attention in this context can be formalized as follows. Given an EEG signal vector $X_t \in \mathbb{R}^{1 \times T}$ ($T$ is the length of the temporal sequence), an attention weight vector $W_t \in \mathbb{R}^{1 \times (n\cdot t)}$ is initialized. Here, $n$ denotes the number of time windows, and $t$ is the length of each window, with $n\cdot t=T$. These parameters can either be fixed as hyperparameters or learned dynamically. The weighted output $\hat{X}_t$ is computed as:
\begin{equation}
\hat{X}_t= \sigma (X_t \odot W_t) 
\end{equation}
where $\sigma$ is an activation function, and $\odot$ denotes element-wise multiplication. This formulation enables the model to focus on critical time segments that are aligned with cognitive task demands.
Furthermore, research in temporal attention for EEG is increasingly oriented toward dynamic, neuroscience-inspired adaptability and data-efficient unsupervised learning. One promising direction involves drawing from neural coding principles, such as the variable temporal resolution observed in biological spiking neural networks, to enable models to adjust temporal granularity in a task-sensitive manner, eliminating the need for heuristic fixed windows. Additionally, unsupervised learning approaches such as contrastive learning or masked segment prediction offer a pathway for discovering generalizable temporal patterns from large volumes of unlabeled EEG data. These methods help mitigate the limitations of scarce annotated datasets while extracting meaningful representations from latent brain dynamics. To further enhance interpretability, generative models can be used to visualize attention-aligned neural features, and causal inference techniques may uncover how specific time segments contribute to cognitive or behavioral outcomes. Together, these innovations offer a pathway toward more adaptive, efficient, and explainable EEG-based systems, with broad implications for clinical neuroscience and next-generation brain-computer interfaces.

\textbf{(3) Frequency Attention Module.} 

The frequency attention mechanism is designed to capture and enhance the spectral characteristics of EEG signals, which reflect brain activity across different cognitive and behavioral states. Frequency-domain analysis has long been recognized as a powerful approach in neuroscience and signal processing. Decades of research in neuroscience and signal processing have shown that specific frequency bands, namely $\delta$ (0.5-4 Hz), $\theta$ (4 - 8 Hz), $\alpha$ (8 - 12 Hz), $\beta$ (13 - 30 Hz), and $\gamma$ ($>$ 30 Hz), are closely associated with distinct brain functions. In practical BCI applications, this predefined division of frequency bands is often used as prior knowledge to guide the analysis of task-related EEG features. The frequency attention module builds on this foundation by learning to adaptively assign higher weights to the most informative spectral components, such as power spectral density (PSD) or differential entropy (DE), thereby improving the extraction of features that are most relevant to tasks like auditory attention detection, motor imagery, and emotional state classification\cite{cai2021eeg, xiao20224d}.

In practical use, the frequency attention mechanism is typically implemented by first decomposing EEG signals into multiple sub-bands based on prior neuroscience knowledge, and then applying attention weights to each band. For example, Cai. \etal decomposed EEG into five classical frequency bands and applied dynamic weighting to identify those most relevant to auditory attention tasks\cite{cai2021eeg}. This method improved detection accuracy compared to models using fixed frequency bands. Similarly, Xie \etal developed a neural model that calculates the importance scores of these frequency bands using convolutional and fully connected layers, offering an efficient solution for real-time decoding with short time windows \cite{xie2024eeg}. Other studies, such as that by Chen \etal \cite{chen2024eeg}, extracted DE features from the five standard bands through manual preprocessing and matched them with attention weights to improve classification performance. A generalized formulation of frequency attention mechanism can be expressed as follows. Given a frequency vector $X_F \in \mathbb{R}^{1 \times F}$, the signal is divided into five frequency band subsets $F^\delta,F^\theta,F^\alpha,F^\beta,F^\gamma$, with each processed into a common feature dimension $f^i$, using a neural network or entropy function:
\begin{equation}
f^i= function (F^i) \qquad i \in \{\delta,\theta,\alpha,\beta,\gamma\}
\end{equation}
The resulting band-wise features are stacked into a matrix $X_f \in \mathbb{R}^{5 \times f}$, followed by the initialization of an attention weight vector $W_f (f \in {\delta,\theta,\alpha,\beta,\gamma})$. The final frequency-aware feature representation is obtained via:
\begin{equation}
\hat{X}_f= \sigma (X_f \odot W_f) 
\end{equation}
This process ensures that the model selectively emphasizes the most relevant spectral components, enhancing its ability to extract robust and discriminative features from EEG data.

Future developments of frequency attention mechanisms may focus on improving adaptability, efficiency, and interpretability. One promising direction is to move beyond static, predefined frequency bands by incorporating data-driven frequency decomposition methods that allow the model to discover meaningful spectral patterns tailored to specific individuals or tasks. Additionally, self-supervised learning techniques such as masked frequency modeling or contrastive learning could enable the model to extract useful frequency features from unlabeled EEG data, thus addressing the common challenge of limited annotations. To enhance interpretability, future work may also integrate generative models that visualize the relationship between attention weights and neurophysiological signals, providing clearer insights into the model's decision-making process. These advancements are expected to improve the flexibility and transparency of EEG analysis, reinforcing the role of frequency attention as a critical component in advanced BCI systems.

The aforementioned mentioned attention modules, including channel, temporal, and frequency, can be used individually or in combination based on the specific demands of the task. The selection and integration of these attention modules are determined by the unique characteristics of the problem being addressed. By strategically choosing the appropriate combination of attention mechanisms, researchers can effectively customize EEG signal processing and model design, providing a more adaptable and efficient solution tailored to the task’s requirements.

\subsection{Embedding Methods of Attention Modules}

\begin{figure*}
\centering
\includegraphics[width=0.5\textwidth]{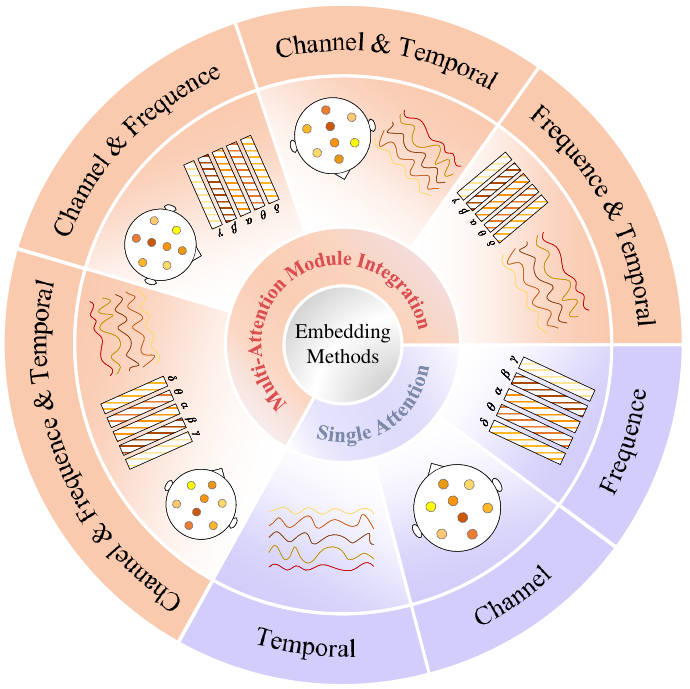}
\caption{Embedding Methods of Attention Modules}
\label{fig:embedding}
\end{figure*}

In embedding attention modules, two primary approaches are employed: single attention module embedding and multi-attention module integration. The single attention module approach thoroughly explores the internal dynamics, efficacy, and performance of a particular attention module across diverse application scenarios, providing insights into how it influences model learning and performance \cite{lv2022attention}. This allows researchers to tailor optimization strategies for specific tasks and datasets. In contrast, the multi-attention module approach integrates multiple attention modules, leveraging their complementary strengths to handle complex data more effectively \cite{fukui2019attention}. This integration enhances the model's generalization capabilities and facilitates a deeper understanding of how different attention mechanisms interact and contribute to information extraction, as illustrated in Fig. \ref{fig:embedding}.

\textbf{(1) Single Attention Module Embedding.} In BCI modeling, embedding a single attention mechanism is widely used approach to enhance a model's ability to focus on critical features. The key idea is to emphasize specific information dimensions (channel or temporal or frequency) by leveraging a particular type of attention mechanism, thereby improving feature extraction and recognition efficiency.

For example, embedding a channel attention module specifically enhances the model’s ability to capture important features at the channel level. Du \etal proposed the ATDD-LSTM model, which combined a channel attention module with a long short-term memory (LSTM) network. In this approach, the channel attention module was applied to feature vectors extracted by the LSTM layers, allowing the model to concentrate on channels most relevant to specific emotions while downplaying less relevant ones, which improved the accuracy of emotion recognition \cite{du2020efficient}. Although channel-level attention mechanisms in EEG analysis have shown promising results, conventional methods often flatten all channels into a uniform space when estimating their importance, disregarding the inherent spatial structure among electrodes, which may introduce bias. To overcome this limitation, Xu \etal integrated channel attention into a graph convolutional network (GCN), taking into account the spatial relationships between EEG recording electrodes \cite{xu2023dagam}. Beyond channel attention, some studies focus on embedding temporal attention module to emphasize the temporal dynamics of EEG signals. Zhang \etal proposed a convolutional recurrent attention model that used CNNs to encode high-level representations and combined them with recurrent attention mechanisms (including LSTM networks and temporal attention modules). This method calculated attention weights on dynamic temporal features, allowed the model to focus on the most informative time periods, and extracted more valuable temporal information \cite{zhang2019convolutional}. Traditional temporal attention modules typically rely on a purely data-driven learning paradigm, where the model autonomously identifies significant temporal segments. However, incorporating prior knowledge to guide the learning process can help the model identify key temporal patterns more efficiently. For example, inspired by the psychological peak-end rule, Kim \etal developed a model that integrated a bidirectional LSTM network with a temporal attention module. It assigned greater weight to emotional states occurring at key moments, capturing the dynamic variability of emotions over time and enhancing the model's interpretability \cite{kim2020eeg}. For frequency-domain features, frequency attention is rarely modeled in isolation. Instead, it is typically integrated with spatial and temporal features to enhance EEG signal representation.

While embedding a single attention module can effectively improve performance in specific tasks, several challenges remain. Choosing the right attention module is crucial. Determining how to integrate it optimally into the model framework for different scenarios is also important. Additionally, understanding how the placement of the attention layer affects model performance requires further exploration. Researchers need to carefully assess both the model architecture and the task requirements to design an optimal embedding strategy.

\textbf{(2) Multi-Attention Module Integration.} Embedding multiple attention modules helps overcome the limitations of single attention module embedding by enabling the model to simultaneously capture various aspects of different feature dimensions. This approach enhances the model’s capacity to learn diverse and informative features, thereby improving its robustness and generalization capabilities. Consequently, transitioning from single to multiple attention embedding is a natural step to better manage the complexity of real-world EEG data.

For example, Tao \etal proposed a deep learning model that incorporated both channel attention and inter-sample attention mechanisms. Since the samples were segmented based on time, the inter-sample attention effectively functioned as temporal attention. This approach enabled the model to effectively prioritize significant information across different channels and temporal segments for feature extraction \cite{tao2020eeg}. Besides of the aforementioned method, Cai \etal introduced a dynamic attention mechanism that assigned different weights to different frequency sub-bands and channels of EEG signals \cite{cai2021eeg}. This dynamic approach optimized feature representation and was applied within an adaptive decoding framework for complex downstream tasks. Additionally, to better capture the spatial relationships among EEG electrode channels, Jia \etal's GraphSleepNet \cite{jia2020graphsleepnet} and Zhang \etal's hierarchical attention network based on graph structures \cite{zhang2019graph} both utilized GCNs to model the spatial relationships of EEG electrodes. These approaches leveraged attention mechanisms across both time and space, significantly enhancing performance in tasks such as sleep stage classification and movement intention prediction. To extract information from EEG signals in a more comprehensive manner, a number of studies have implemented attention mechanisms across all three dimensions (channel, temporal and frequency). However, expanding attention to multiple dimensions requires careful architectural design to ensure a balanced and effective focus across them. Extending beyond channel and temporal information, Jia \etal introduced a spatio-temporal-spectral attention dense network that simultaneously considered temporal, frequency, and spatial features. This model adaptively captured crucial information across brain regions (spatial, i.e., channels), frequency bands (spectral), and time, providing a comprehensive feature extraction framework \cite{jia2020sst}. Xiao \etal extended Jia et al.'s model by proposing a neural network based on 4D attention \cite{xiao20224d}. In this approach, the channel dimension of the input samples is transformed into a two-dimensional feature to preserve the spatial positional information of the EEG signal electrodes, while also incorporating the time and frequency dimensions. It computed spatial attention (addressing the spatial positional relationships between channels) and frequency attention (applied to power spectral density and differential entropy features). These attention weights were then applied to refine the input, resulting in enhanced output features. Jia \etal's and Xiao \etal's models both leveraged temporal, frequency, and spatial characteristics of EEG channels, calculating attention weights across these dimensions to enhance the model’s focus on information pertinent to specific tasks. The main difference between their approaches lies in how they calculated attention weights and structured their models. Jia \etal computed attention separately across the frequency and time dimensions in parallel, and then merged these features for classification. In contrast, Xiao \etal integrated all dimensions into a unified 4D representation before computing attention, providing a different approach to capture feature interdependencies. While Jia \etal’s dual-stream framework excels in modular feature extraction, it may miss cross-dimensional dependencies. In comparison, Xiao \etal’s integrated 4D strategy enables holistic feature learning but introduces greater model complexity. The choice between these approaches depends on the dependency structure of the task: Jia \etal’s method is more suited for problems where temporal and spectral features are relatively independent, whereas Xiao \etal’s model is preferable when multi-dimensional interactions are crucial.

 A crucial aspect of applying attention mechanisms is defining the shape and dimensions of the input feature matrix. In previous research, attention weight computations have predominantly relied on the strict Euclidean geometric space of the feature matrix. However, given the brain’s complex topological structure, using Euclidean space alone may not accurately capture its underlying properties. Consequently, several studies have sought to align feature matrix definitions more precisely with the brain’s physiological structure by incorporating non-Euclidean space representations within attention mechanisms. For example, Zhang \etal introduced the concept of manifolds, proposing a time-frequency domain feature learning model that integrated both Riemannian manifold and Euclidean space representations. Their work demonstrated the effectiveness of attention mechanisms in synthesizing feature information across different mathematical domains \cite{zhang2020spatio}. 

These studies collectively highlight the crucial role of attention mechanisms in enhancing the performance of EEG signal processing models, especially in terms of accuracy and efficiency of feature extraction. By calculating attention weights across multiple dimensions, such as channel, temporal, and frequency, and either integrating or applying them independently, these models offer more refined and effective solutions to address the complexities of EEG signal processing.The relevant literature we have reviewed is summarized in Table \ref{tab:table11}.

\begin{table*}[]
\centering
\caption{Embedding methods of attention modules in the literature.}
\label{tab:table11}
\resizebox{1\textwidth}{!}{
\begin{tabular}{lccccc}
\hline
Ref  & Year & Embedding style  & Backbone & Task  & Dataset   \\ \hline
\cite{zhang2019graph} & 2019 & Channel, Temporal & CNN, GCN  & Movement decoding & PhysioNet\\
\cite{zhang2019convolutional} &2019 &Temporal &CNN, LSTM & Movement intention recognition & BCI IV-2A \\
\cite{du2020efficient} & 2020 & Channel & LSTM & Emotion recognition & DEAP, SEED, CMEED \\
\cite{jia2020graphsleepnet} & 2020 &Channel, Temporal & GCN  & Sleep stage recognition  & MASS-SS \\
\cite{tao2020eeg} & 2020 & Channel, Temporal & RNN & Emotion recognition & DEAP, DREAMER         \\
\cite{kim2020eeg}   &2020 &Temporal  &LSTM, CNN   &Emotion recognition  &DEAP \\
\cite{jia2020sst}  & 2020 & Channel, Temporal, Frequency & CNN & Emotion recognition & SEED, SEEDIV \\
\cite{zhang2020spatio} & 2020 & Temporal, Frequency & LSTM,CNN & Emotion recognition & \makecell[c]{SEED-VIG, SEED,\\ BCI IV-2A, BCI IV-2B} \\
\cite{su2021auditory} & 2021 & Channel & CNN & Auditory attention detection & KUL \\
\cite{cai2021eeg} & 2021 & Channel, Frequency & CNN & Auditory attention detection & KUL, DTU \\
\cite{chen2021multiattention} & 2021 & Channel, Frequency & CNN & Motor imagery & BCI IV-2A, BCI IV-2B, HGD \\
\cite{xiao20224d}  & 2022 & Channel, Temporal, Frequency & LSTM, CNN & Emotion recognition & DEAP, SEED, SEEDIV   \\
\cite{wu2023classification} & 2023 & Channel, Temporal & CNN & Motor imagery & BCI IV-2A, Custom \\
\cite{xu2023dagam} & 2023 &  Channel &GCN & Emotion recognition & SEED, SEEDIV  \\
\cite{hsu2023eeg} & 2023 &  Channel &CNN & Motor imagery & BCI IV-2A,BCI IV-2B, 2020BCI Track\#1 \\
{\cite{wimpff2024eeg}} & {2024} & {Channel} & {CNN} & {Motor imagery} & {BCI IV-2A, BCI IV-2B, HGD, BCI III-4A} \\
{\cite{yuan2024psaeegnet}} & {2024} & {Channel, Temporal} & {CNN} & {RSVP} & {Tsinghua RSVP} \\
{\cite{wu2025ameegnet}} & {2025} &  {Channel} & {CNN} & {Motor imagery} & {BCI IV-2A, BCI IV-2B, HGD} \\
{\cite{liao2025composite}} & {2025} &  {Temporal} & {CNN} &   {Motor imagery} & {BCI IV-2A, BCI IV-2B} \\
\hline
\end{tabular}
}
\end{table*}

\section{Transformer-based Multi-Head Self-Attention Mechanisms in EEG}

\begin{table*}[]
\centering
\caption{Transformer-based embedding methods in the literature.}
\label{tab:table12}
\resizebox{1\textwidth}{!}{
\begin{tabular}{lccccc}
\hline
Ref & Year & Embedding style  & Backbone & Task   & Dataset \\ \hline
\cite{qu2020residual}  & 2020 & Temporal & Transformer & Sleep stage recognition & MASS, Sleep-EDF \\
\cite{song2021transformer} & 2021 & Temporal, Channel & Transformer & Motor imagery & BCI IV-2A, BCI IV-2B \\
\cite{li2022eeg} & 2022 & Frequncy, Temporal & Transformer & Seizure prediction  & CHB-MIT, Kaggle datasets \\
\cite{guo2022transformer} & 2022 & Channel & Transformer & Emotion recognition & SEED \\
\cite{zheng2022copula}  & 2022 & Temporal, Channel & Transformer & Visual discomfort induces & Private datasets \\
\cite{du2022eeg} & 2022 & Temporal, Channel & Transformer & Motor imagery & PhysioNet \\
\cite{si2024temporal}  & 2024 & Channel, Temporal & Transformer  & Emotion recognition & {DEAP, SEED, THU-EP} \\ 
{\cite{pan2024dua}}  & {2024} & {Channel, Frequncy} & {Transformer}  & {Emotion recognition} & {Private datasets, SEED, SEED-IV} \\ 
{\cite{li2024dual}}  & {2024} & {Temporal, Channel, Frequncy} & {Transformer}  & {Emotion recognition, Motor imagery} & {BCI IV-2A, BCI IV-2B, SEED} \\ 
{\cite{ding2025decoding}}  & {2025} & {Temporal, Channel} & {Transformer}  & {Cognitive attention classification} & {A publicly accessible EEG dataset} \\ 
{\cite{pradeepkumar2025single}}  & {2025} & {Temporal} & {Transformer}  & {EEG event recognition, epilepsy seizure detection} & {TUEV, TUAB, IIIC Seizure, CHB-MIT} \\
\hline
\end{tabular}}
\end{table*}

The Transformer model, first proposed by Vaswani \etal in 2017 \cite{vaswani2017attention}, introduced a self-attention mechanism that revolutionized machine translation tasks. The Transformer architecture is composed of encoder and decoder modules, which are built by stacking multiple sub-layers, including self-attention, feed-forward neural networks, residual connections, and normalization layers \cite{vaswani2017attention,devlin2018bert}. This composition enables the model to effectively encode input sequences and generate outputs, with the self-attention mechanism at its core enhancing the ability to capture contextual and temporal relationships \cite{kobayashi2020attention}.

Since then, the Transformer and its variants have seen widespread application across fields such as natural language processing, and computer vision \cite{lin2022survey}. The core strength of the Transformer lies in its capacity to capture long-range dependencies and interactions among input features, making it particularly effective for time series modeling and also achieving significant advancements in temporal signal processing task. Table \ref{tab:table12} summarizes a survey of studies that employ the transformer architecture. For example, AST \cite{wu2020adversarial} utilized a generative adversarial encoder-decoder framework to train a sparse Transformer model for time series prediction. It demonstrates that adversarial training can enhance time series prediction by directly shaping the network's output distribution to mitigate error accumulation during one-step-ahead inference. FEDformer \cite{zhou2022fedformer} applied attention mechanisms in the frequency domain using Fourier and wavelet transforms, achieving linear complexity by randomly selecting a fixed-size subset of frequencies. It is noted that FEDformer's success has spurred increased interest in exploring the self-attention mechanism in the frequency domain for time series modeling \cite{wen2022transformers}.

\subsection{Multi-Head Self-Attention Mechanisms}

The self-attention mechanism, also known as "Scaled Dot-Product Attention" \cite{xu2023multimodal}, offers a significant advantage over traditional attention models. It captures contextual relationships effectively, especially in long sequences. This helps overcome challenges like information loss and long-term dependencies. Self-attention analyzes correlations between positions in the input sequence, making it more efficient than CNNs and RNNs for sequence modeling.

In the Transformer model, self-attention is implemented using three matrices: Query ($Q$), Key ($K$), and Value ($V$). These matrices are derived from the input feature matrix $I \in \mathbb{R}^{L \times D}$, where $L$ is the sequence length and $D$ is the feature dimension. The matrices $Q$, $K$, and $V$ are obtained by applying linear transformations to $I$:

\begin{equation}
Q = I \otimes W^Q,
\end{equation}
\begin{equation}
K = I \otimes W^K,
\end{equation}
\begin{equation}
V = I \otimes W^V,
\end{equation}
where $W^Q \in \mathbb{R}^{{D \times D_k}}$, $W^K \in \mathbb{R}^{{D \times D_k}}$, and $W^V \in \mathbb{R}^{{D \times D_v}}$ are trainable weight matrices. The self-attention layer then calculates attention weights and the output:
\begin{equation}
A = \text{softmax}(\frac{Q K^T}{\sqrt{D_k}}).
\end{equation}
\begin{equation}
\text{Attention}(Q,K,V) = A  V.
\end{equation}

To allow multiple self-attention processes to run in parallel, the multi-head attention mechanisms is suggested. For $H$ self-attention heads, the output of the multi-head attention is:
\begin{equation}
\text{MultiHeadAttn}(Q,K,V) = \text{Concat}(head_1,\ldots,head_H)W^O,
\end{equation}
where $W^O$ is a trainable weight matrix. This ensures the output size matches the input. The multi-head approach allows the model to focus on different parts of the sequence simultaneously. This enriches the representation and improves model performance.

\begin{figure*}[h]
	\centering
	  \includegraphics[width=1\textwidth]{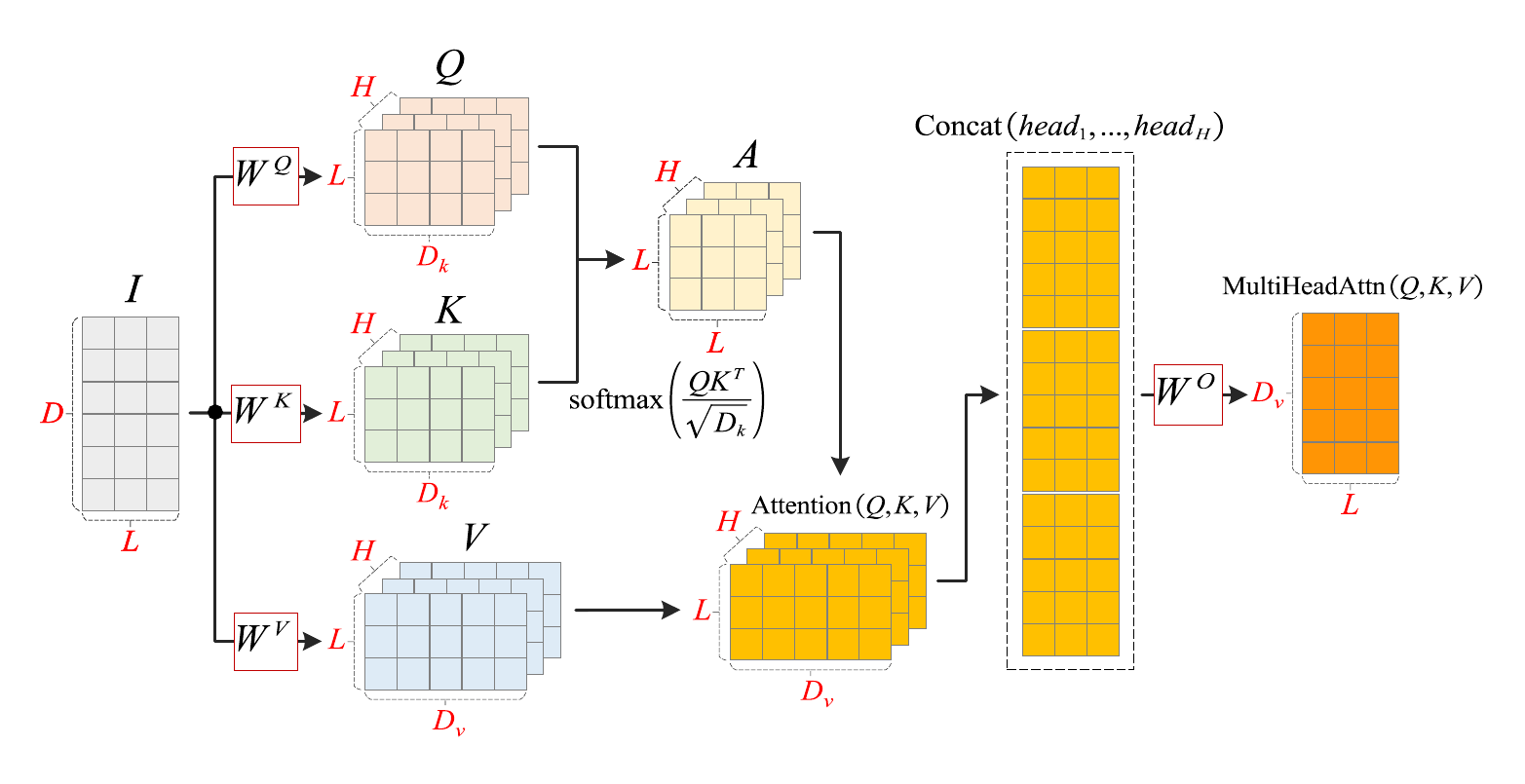}
	\caption{Multi-Head Self-Attention Module}
 \label{fig:multi-head}
\end{figure*}

\subsection{Strategies for Applying Transformers}

In practical applications, the complete Transformer architecture is not always necessary for every task. Many models modify specific Transformer components or integrate elements into existing frameworks. Broadly, Transformer applications can be categorized into three main strategies \cite{lin2022survey}.

\textbf{(1) Encoder-Decoder Combination.} This strategy suits sequence-to-sequence tasks, where the input sequence is processed by an encoder and then decoded into a target sequence. This approach addresses long-range dependencies by fully leveraging contextual information. While this method increases model complexity and requires more parameters, longer training times, and larger datasets, it generally yields improved performance.

\textbf{(2) Encoder Only.} Used for non-sequence-to-sequence tasks, this strategy employs only the encoder to convert the input sequence into a specialized representation for subsequent processing. By simplifying the model structure and reducing parameters and training time, it provides a more efficient approach. However, the encoded representation may lack the depth of contextual information needed for complex sequence generation tasks.

\textbf{(3) Decoder Only.} Typically paired with a pre-trained encoder module, this strategy is ideal for generative tasks. The decoder, utilizing self-attention, generates the target sequence based on the representations from upstream tasks. This setup captures comprehensive contextual information, though it may increase time complexity.

In BCI applications, the core benefit of Transformers lies in the self-attention mechanism, essential for capturing temporal correlations and performing effective feature encoding. Consequently, BCI applications frequently implement either the encoder or encoder-decoder strategy, with self-attention as a pivotal component for handling EEG-based tasks.

\subsection{Practical Applications of Transformer Models in EEG Analysis}
The use of Transformer-based self-attention mechanisms shows great potential for enhancing EEG modeling. These mechanisms help capture relevant information from complex, non-stationary EEG signals. We introduce EEG modeling with Transformers in three areas: temporal, spatial, and combined temporal-spatial.

\textbf{(1) Application in Temporal Dimension.}
In the temporal domain, researchers have increasingly adopted hybrid models such as CNN-Transformer and LSTM-Transformer to extract informative temporal representations from EEG signals \cite{xie2022transformer}. Enhancements to the self-attention mechanism have significantly improved the models’ ability to capture evolving neural patterns and temporal dynamics. Qu \etal \cite{qu2020residual} proposed a residual-based attention model for sleep staging, treating 30-second EEG epochs as fundamental processing units. The model’s decoder utilized a multi-head self-attention mechanism, enriched with sinusoidal positional encoding, to retain the sequential structure of sleep transitions. By embedding residual blocks within the encoder for feature extraction and adopting parallelizable self-attention operations, the model effectively captured inter-epoch dependencies (e.g., sleep stage shifts) while mitigating the sequential processing limitations inherent in LSTM architectures. This design achieved a 12× acceleration in training speed, enabling efficient large-scale sleep data processing. Additionally, the model performed multi-scale frequency decomposition using Hilbert-transform-inspired pooling, and directly operated on raw EEG inputs in an end-to-end architecture, thus eliminating reliance on handcrafted preprocessing. Complementing this work, Li \etal \cite{li2022eeg} introduced the Transformer-guided CNN (TGCNN) for epilepsy prediction, featuring a novel squeezed multi-head self-attention (SMHSA) module. Unlike standard Transformer attention, SMHSA applied depth-wise convolutions to downsample the Key and Value matrices, reducing computational cost by a factor of four, while incorporating learnable bias terms to improve positional encoding. The model employed an alternating CNN-Transformer structure, in which CNN layers first extracted local time-frequency patterns from STFT-based spectrograms, followed by SMHSA modeling of long-range temporal dependencies. This allowed the model to capture both fine-grained waveform details (e.g., seizure spikes) and broader temporal trends (e.g., pre-ictal changes), achieving a robust balance between local feature resolution and global contextual awareness. While Qu \etal focused on optimizing training efficiency and temporal transition modeling for regular and stable sleep stages, Li \etal targeted computational scalability and feature complementarity to detect rare, transient epileptic events. However, each approach has its trade-offs: Qu's reliance on fixed-length epochs may limit modeling of ultra-long temporal dependencies, whereas Li’s use of STFT preprocessing could result in loss of raw phase information. Together, these studies underscore the importance of task-specific adaptations of self-attention (such as residual-enhanced encoding in Qu’s model and compressed attention in Li’s architecture) in optimizing EEG temporal modeling. Future research directions may include the integration of adaptive temporal windows, multi-resolution temporal attention, or self-supervised pretraining to improve generalization across diverse EEG analysis tasks.

\textbf{(2) Application in Spatial Dimension.}
Inspired by the Vision Transformer (ViT) model for image processing, researchers have developed innovative neural network models tailored for spatial feature analysis in EEG signals. For example, Guo \etal proposed a model called the Deep Convolutional and Transformer Network (DCoT) \cite{guo2022transformer}, focusing on the importance of each EEG channel in emotion recognition and visualizing these features. Based on the extracted differential entropy features, EEG signals were formed in a three-dimensional representation (Time $\times$ Channel $\times$ Frequency), which is similar to RGB representation in image. Inspired from ViT structure, DCoT was designed to analyze correlations between EEG electrodes, with each Transformer token representing a specific EEG electrode channel. The model applied positional encoding to the input tokens (i.e., EEG channels) and introduced an additional token dedicated to classification. After processing the encoded signals through the Transformer encoder module, a fully connected layer generated the final classification results. Furthermore, Zheng \etal proposed a Copula-Based Transformer model (CBT) \cite{zheng2022copula}, designed to learn spatial dependencies between EEG channels while optimizing classification performance. By reducing the size of the attention matrix, CBT lowered dependency on large datasets, improving computational efficiency. The CBT model excelled in an EEG-based visual discomfort assessment task, pioneering the revelation of temporal characteristics in EEG signals linked to visual discomfort. These studies underscore that Transformer-based self-attention mechanisms effectively capture the spatial dynamics of EEG signal sequences, enhancing spatial information processing and optimizing model training efficiency and accuracy.
These studies illustrate two complementary strategies for integrating spatial information into Transformer-based EEG models. DCoT adopts a data-driven self-attention paradigm, enabling the dynamic discovery of inter-channel correlations, and is particularly well-suited to applications demanding interpretability, such as emotion recognition with spatial brain mapping. In contrast, CBT leverages structured priors via copula functions to explicitly encode electrode positioning, which is advantageous in scenarios characterized by limited data availability and region-specific neural responses, such as visual discomfort detection. While both approaches validate the effectiveness of self-attention for spatial feature modeling, future research could explore hybrid frameworks that combine physiological priors (e.g., brain connectivity graphs, cortical maps) with adaptive attention mechanisms. Such integration may further enhance the generalization capability of EEG-based models across diverse cognitive and clinical tasks, especially when spatial specificity is critical.

\textbf{(3) Application in Combined Temporal and Spatial Dimensions.}

Researchers have leveraged the powerful capabilities of the Transformer self-attention mechanism to simultaneously capture both the temporal and spatial dimensions of EEG signals, enabling the extraction of highly discriminative features. For example, Song \etal \cite{song2021transformer} developed an EEG decoding model that primarily relies on the Transformer’s self-attention layers to enhance feature representations in both dimensions. In the spatial transformation component, the self-attention mechanism weighted each channel, emphasizing signals with higher relevance. Meanwhile, the temporal transformation component employed convolutional layers to encode temporal features, with channel compression and sample segmentation steps to reduce computational load. Finally, global average pooling and fully connected layers were combined to classify EEG signals effectively. Similarly, Du \etal proposed an EEG spatio-temporal Transformer network \cite{du2022eeg}, which introduced Temporal Transformer Encoder (TTE) and Spatial Transformer Encoder (STE) modules to independently capture temporal and spatial features. In the TTE, temporal attention mechanisms calculated correlations among sampling points within each sample, extracting temporal features. Since channel correlations are often unique to individuals, the STE calculated spatial attention among channels and applies positional encoding to retain spatial location information. This allows the model to capture inter-channel relationships more accurately, improving its ability to identify individuals based on unique EEG patterns. Additionally, Si \etal \cite{si2024temporal} proposed integrating the Selective Kernel (SK) attention mechanism with the Transformer self-attention mechanism to identify and select the channels most relevant to the current task. This approach enables the model to filter out noise and irrelevant signals, simplifying the processing of complex temporal information. Once the SK attention mechanism isolated key channels, the Transformer encoder module deeply extracted temporal features from these selected channels. This approach's strength lies in its ability to focus on the most informative parts of the data while preserving sensitivity to temporal dependencies, a critical factor for managing complex tasks in BCIs. These studies demonstrate that an incorporation of both temporal and spatial information can significantly enhance the accuracy and efficiency of EEG signal analysis. A critical factor for managing the variability and complexity of real-world BCI scenarios. Together, these studies illustrate the advantages of incorporating both temporal and spatial information within Transformer-based EEG models. Each approach presents task-specific strengths. Si \etal enable robust emotion recognition via hierarchical attention. Song \etal achieve efficient motor imagery decoding with lightweight slice-based self-attention. Du \etal enhance person identification using spatio-temporal encoders and positional encoding. {These works highlight the growing trend of leveraging modular attention architectures to tailor EEG decoding strategies for diverse BCI applications.}

\section{Attention Models for Multimodal Applications}
\begin{figure}[h]
	\centering
	  \includegraphics[width=0.45\textwidth]{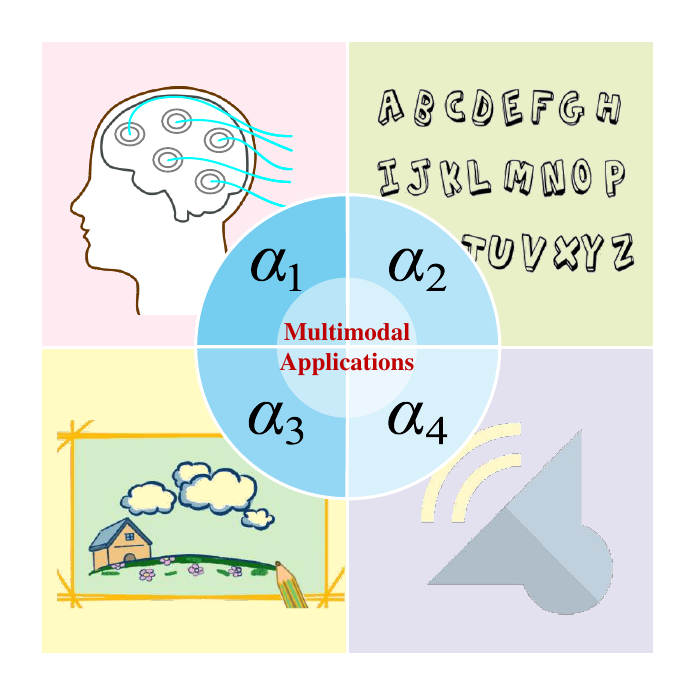}
	\caption{Attention Modules for Multimodal Applications}
 \label{fig:multi-model}
\end{figure}

To improve the accuracy of EEG signal recognition, recent studies have gradually introduced multimodal data (such as speech, images, text, etc.) to be jointly trained with EEG signals (as shown in Fig. \ref{fig:multi-model}). By leveraging traditional attention mechanism modeling strategies or Transformer-based multi-head self-attention mechanism modeling strategies, these approaches achieve effective fusion of signals from different modalities, enhancing the accuracy and stability of recognition. As multimodal applications in brain modeling continue to expand, the ability to efficiently utilize information from each modality becomes increasingly critical.

At the current stage, multimodal tasks face two core challenges: (1) effectively fusing multimodal data and (2) facilitating better interaction between different modalities. The key to the first challenge lies in assessing the importance of each modality for a given task and assigning appropriate weights to generate a unified feature representation for downstream processing. The selection of these weights significantly impacts model performance. Traditionally, this process relies on manual tuning, which is both time-consuming and suboptimal. In contrast, automated weight optimization not only improves efficiency but also enhances model performance. This is typically achieved by incorporating attention models that introduce an attention parameter matrix to dynamically adjust weight distribution. Attention modeling can be broadly categorized into traditional attention mechanisms and Transformer-based self-attention strategies. The latter emphasizes capturing complementary and shared information between modalities, ensuring that the information from one modality is reflected in another. For example, in cross-modal generation tasks using diffusion models, cross-attention mechanisms are often employed to facilitate information exchange and allocate importance between modalities.

The second challenge focuses on improving interactions between different modalities, ensuring seamless information exchange across data sources. A major difficulty arises from the heterogeneous nature of multimodal data, where differences in feature distributions, temporal alignment, and semantic relevance must be addressed. To tackle this, recent studies have introduced contrastive learning \cite{radford2021learning}, cross-modal attention \cite{rombach2022high}, and graph-based approaches \cite{ding2023mst} to establish meaningful associations between modalities. These methods enhance feature complementarity and reduce modality gaps, ultimately improving model robustness and generalization. Furthermore, explicit modeling of inter-modality dependencies through techniques such as mutual information maximization \cite{cao2024predictive} and self-supervised learning strengthens interactions \cite{wei2023multi}, allowing each modality to contribute more effectively to the overall task performance.

\subsection{Application of Attention Mechanisms in the Fusion of EEG Signals with Multimodal Data}

In multimodal emotion recognition tasks, Liu \etal employed two modality fusion strategies: a basic weighted fusion method and an attention-based fusion method \cite{liu2021comparing}. In the basic weighted fusion approach, the weight coefficients for each modality were manually adjusted, and the weighted sum of data from different modalities was computed to generate a fused output. The fusion process was represented as:
\begin{equation}
     O = \alpha_1O_1+\alpha_2O_2.
\end{equation}
Here, $\alpha_1$ and $\alpha_2$ represented the weight coefficients assigned to different modalities, satisfying $\alpha_1 + \alpha_2 = 1$. The model’s performance was thoroughly evaluated by manually testing various combinations of $\alpha_1$ and $\alpha_2$. In the attention-based fusion method, although a weighted summation formula was still used to obtain the fused feature $O$, in solving for $\alpha_1$ and $\alpha_2$, the authors first initialized an attention layer, which was then optimized through the network. The final attention weights, $\alpha_1$ and $\alpha_2$, were derived from the softmax function applied to the attention layer’s output. This approach allows the model to adaptively adjust the attention weights, dynamically learning an optimal set of weight parameters.
This work attempts to enhance multimodal fusion by incorporating adaptive weight coefficients. It systematically compares traditional manual weighting with an attention-based dynamic adjustment strategy, ultimately demonstrating improvements in both performance and robustness for multimodal emotion recognition. These findings further support the application potential of multimodal analysis in BCI systems. However, the model architecture was restricted to the fusion of only two modalities, and the employed attention mechanism was relatively simple in design. On the other hand, Qiu \etal designed a more complex attention structure. They developed a correlation attention network that effectively integrated eye movement and EEG signals for fusion analysis. The unique aspect of the correlation attention network is that it not only computed attention weights for the two modalities but also adopted the canonical correlation analysis (CCA) method from statistics. This analysis method generated a cross-correlation coefficient matrix $C$ by calculating the cross-correlation between outputs of each recurrent unit in the bidirectional gated recurrent unit (GRU) layer. Utilizing this matrix along with the average features extracted by the bidirectional GRU, the fusion feature matrix $F$ was constructed to determine the attention weights. The computation followed a structured process. The attention score $u_{it}$ at each time step was derived as
\begin{equation}
     u_{it} = tanh(W_{w1}f_{it}+W_{w2}c_{it}+b_w)
\end{equation}
The attention weights for each time point were obtained by normalizing the attention scores using the softmax function $\alpha_{it}$, as
\begin{equation}
     \alpha_{it} = \frac{\exp{(u_{it}^T})}{\sum_{t} \exp{(u_{it}^T})}
\end{equation}
The final attention-weighted output $s_i$ was computed as follows:
\begin{equation}
     s_i = \sum_{t}\alpha_{it}h_{it}
\end{equation}
Here, $f_{it}$ and $c_{it}$ represented elements from the fusion feature matrix $F$ and the cross-correlation coefficient matrix $C$, respectively. $W_{w1}$, $W_{w2}$, and $b_w$ were trainable weight parameters. This approach effectively combined the complementarity of multimodal data with statistical correlation analysis, introducing an innovative attention mechanism for multimodal tasks. This not only enhances the model's efficiency in understanding and utilizing multimodal data features but also demonstrates how traditional attention mechanisms can be extended and improved through statistical analysis methods.

Recent studies have pointed out that traditional multimodal fusion approaches tend to focus solely on integrating various types of physiological signal features. A novel and increasingly promising direction involves combining human statistical data (such as demographic information) with physiological signals from the brain. This integration provides a richer and more holistic understanding of brain activity and is expected to significantly advance the performance and applicability of BCI systems. Zhang \etal proposed a multimodal neural network model for integrating demographic data and EEG signals \cite{zhang2020eeg}. This model incorporated an attention mechanism to effectively combine the two data types, aiming to uncover complex relationships between EEG signals and demographic factors. Such integration is particularly valuable for applications like depression detection, where demographic information can provide crucial contextual insights. The model first used a one-dimensional CNN to process EEG signals, resulting in a feature matrix $Z \in \mathbb{R}^{C_{out} \times m}$, while demographic data were encoded as a feature vector $S \in \mathbb{R}^d$. The two data types were fused through an attention mechanism, with the calculation as follows:
\begin{equation}
     A = tanh((W_{fe}Z+b_{fe})\oplus(W_{de}S+b_{de})).
\end{equation}
Here, $W_{fe}$ and $b_{fe}$ were trainable weights and biases related to EEG signal features. $W_{de}$ and $b_{de}$ were weights and biases related to demographic features. The symbol $\oplus$ represented the fusion operation between the two modalities. This approach enhances the integration of demographic data and EEG signals by leveraging attention mechanisms, helping to mitigate the impact of individual differences on model performance. By dynamically adjusting the influence of demographic information, the model achieves a more effective and personalized feature representation, improving the accuracy of multimodal analysis.

Furthermore, Choi \etal proposed a multimodal attention network to explore the fusion of facial video and EEG signals \cite{choi2020multimodal}. Unlike traditional methods that rely on a single fusion layer, this network utilized multimodal fusion layers incorporating bilinear and trilinear pooling to extract and integrate deep features. This approach not only enhances the integration of features from both modalities but also improves model performance by dynamically allocating attention weights, enabling more effective cross-modal information exchange.

Recently, Transformer-based multi-head self-attention mechanisms have become a prevalent approach in multimodal data fusion, enabling efficient processing and integration of features from diverse modalities. In the model framework, each token received by the Transformer encoder corresponds to a specific modality, allowing the model to effectively capture and integrate information at the feature level. By leveraging complex inter-modal interactions, the self-attention mechanism uncovers intrinsic relationships and complementary information between modalities, dynamically refining feature representations. A well-known example is the MMASleepNet model \cite{yubo2022mmasleepnet}, which not only incorporated the concept of the Squeeze-and-Excitation (SE) module from SENet \cite{hu2018squeeze} but also integrated the advantages of the Transformer encoder module. To enhance the utilization of multimodal collaborative information, researchers typically employ a two-stage attention mechanism: first, applying attention modules to extract modality-specific features from individual modalities, followed by integrating cross-modal information through another layer of attention during fusion. This strategy, which combines hierarchical attention mechanisms, has demonstrated remarkable effectiveness in BCI tasks. For instance, He \etal proposed the EEG-EMG FAConformer framework\cite{he2024eeg}, which leverages self-attention modules to hierarchically extract critical features from EEG and EMG(electromyography) signals. By further incorporating cross-modal attention for fusion, this approach achieves robust performance and enhanced classification accuracy, as evidenced by its superior stability and generalization capabilities in experimental evaluations. Another interesting research work by Wang \etal  proposed a cross-modal fusion model to combine EEG signals with textual data for enhancing emotion detection in English writing\cite{wang2025cross}. The model utilizes the self-attention mechanism of the Transformer architecture to dynamically fuse EEG signals and textual features, addressing the limitations of traditional methods in leveraging the complementary information between textual and physiological signals.

\subsection{Application of Attention Mechanisms in Information Interaction Between EEG Signals and Multimodal Data}

Compared to self-attention, cross-attention extends its functionality by integrating information from multiple modalities, enabling more precise modeling of inter-modal associations \cite{li2024multimodal}. When the query (Q) matrix and the key (K) and value (V) matrices originate from different modalities, the computation shifts from capturing intra-modal correlations to establishing inter-modal relationships, effectively transforming self-attention into cross-attention \cite{lee2022cross}. Also referred to as cross-modal attention, this mechanism introduces additional input sequences from distinct modalities, enhancing information fusion and improving the overall representation of multimodal data \cite{wang2022scanet}\cite{zhao2024deep}.

In EEG-related research, visual reconstruction from EEG signals is a particularly challenging yet rapidly evolving field. The objective is to decode the information embedded in EEG signals and use it to reconstruct corresponding visual stimuli. This process requires capturing the intricate relationships between EEG and visual modalities, often achieved through cross-attention mechanisms. By computing the relevance between EEG segments and visual data, EEG signals serve as conditioning inputs to guide the generation of visual stimuli. A well-known example is the DreamDiffusion model \cite{bai2023dreamdiffusion}, which employed an EEG encoder trained on large-scale EEG data using a masking strategy to enhance feature extraction. The extracted EEG features then conditioned the Stable Diffusion model to generate images. Within Stable Diffusion, the cross-attention mechanism computes correlations between EEG-derived and image-derived features, enabling coherent information exchange between the two modalities. {The cross-attention mechanism employed in Stable Diffusion model has been widely adopted in EEG-to-image guided reconstruction tasks, such as in works like\cite{li2024realmind,li2024visual}}. The mathematical formulation of cross-attention can be expressed as follows:
\begin{equation}
Q=W_{Q}^{(i)}\cdot\varphi_{i}\left(z_{t}\right),K=W_{K}^{(i)}\cdot\tau_{\theta}(y),V=W_{V}^{(i)}\cdot\tau_{\theta}(y).
\end{equation}
Here, $\varphi_i\left(z_t\right) \in \mathbb{R}^{N \times d_e^i}$ is the output of the noise prediction model Unet, and $\tau_{\theta}(y) \in \mathbb{R}^{M \times d_{\tau}}$ is the encoded representation obtained by projecting the EEG features output from the EEG encoder through an additional layer. $W_{V}^{(i)} \in \mathbb{R}^{d \times d_{\epsilon}^{i}}, W_{Q}^{(i)} \in \mathbb{R}^{d \times d_{\tau}}, \mathrm{and} , W_{K}^{(i)} \in \mathbb{R}^{d \times d_{\tau}}$ are learnable parameters. In this setup, the Q matrix is derived from image data, while the K and V matrices are derived from EEG signals. By substituting these Q, K, and V matrices into the self-attention computation formula, the attention score matrix is obtained, leading to the output of the cross-attention mechanism. This process facilitates effective information interaction between EEG and image modalities, enabling more precise multimodal feature integration. 

\subsection{{Comparison of Transformer-Based Cross-Attention and Traditional Attention in EEG Multimodal Applications}}
{Traditional attention mechanisms and Transformer-based cross-attention have both been applied in EEG multimodal fusion tasks. In recent years, Transformer-based approaches have increasingly become the mainstream. Traditional attention retains limited advantages in computational efficiency and parameter compactness, often using fewer parameters, which suits real-time BCI and low-resource clinical applications\cite{s25051293}. However, Transformer-based models demonstrate superior overall performance. For example, the Dual-Branch Transformer with Cross-Attention (DTCA) achieves state-of-the-art results on SEED and SEED-IV datasets by encoding EEG and eye movement data with sophisticated query-key-value relationships\cite{10.1007/978-3-031-72069-7_14}. In more complex scenarios involving a greater number of modalities, Transformer-based cross-attention mechanisms also demonstrate strong performance. For example, Ying \etal developed a method that integrates EEG, audio, and video, advancing the incorporation of EEG into more complex multimodal systems\cite{yin2025eeg}. Notably, researchers are actively exploring lightweight Transformer architectures. In the future, significant improvements in the computational efficiency of Transformer-based cross-attention mechanisms are expected to further reduce the reliance on traditional attention in multimodal EEG systems.}

\section{Conclusion and Future Works}
This paper presents a systematic and comprehensive review of recent advancements in attention mechanisms integrated into EEG-based BCI systems. We structure the review around three conceptually related categories of attention mechanisms: traditional attention methods, Transformer-based multi-head self-attention architectures, and multimodal attention fusion strategies. First, we introduce traditional attention mechanisms, including channel-specific, temporal-specific, and frequency-specific modules. These approaches enhance EEG signal decoding by adaptively emphasizing salient features, which could improve model accuracy and robustness against noise and inter-subject variability. We then explore Transformer-based architectures that utilize multi-head self-attention mechanisms to capture complex temporal and spatial dependencies within EEG data. These methods enable more expressive and flexible feature representations, addressing limitations of earlier approaches. We then review multimodal attention fusion strategies, which incorporate EEG signals with additional physiological and sensory modalities. This integration enhances system interpretability and adaptability, supporting more effective performance in complex and dynamic environments. Together, these three categories form a coherent taxonomy that reflects the methodological evolution of attention-based techniques in neural decoding and offers a comparative understanding of their respective advantages, limitations, and suitable application contexts. Through this structured overview, we aim to provide researchers with a practical reference for selecting attention mechanisms that best address specific challenges in BCI design.

Despite the progress outlined in this review, several important challenges remain. Current attention mechanisms often suffer from limited generalizability, as their performance tends to be constrained by dataset-specific and subject-specific characteristics. In addition, the high computational complexity of models such as Transformers can hinder their application in real-time and resource-limited scenarios. To address these issues, future research should focus on the development of lightweight, energy-efficient, and EEG-specific attention models that are better aligned with the physiological properties of neural signals. Advancing in this direction will help unlock the broader potential of attention-based BCIs, enabling their practical use across a range of domains including clinical diagnosis, neurorehabilitation, education, and human–machine interaction.

To establish a stronger foundation for future advancements, it is essential to examine the current landscape of attention-based BCI applications across various domains. In clinical neuroscience, attention mechanisms have significantly enhanced diagnostic accuracy in EEG-based emotion recognition and the detection of neurological disorders such as epilepsy and Parkinson’s disease. However, these clinical systems continue to face challenges, particularly with inter-subject variability and the need for extensive calibration, which hinder widespread clinical deployment. In motor rehabilitation, attention mechanisms optimized for specific EEG channels and temporal dynamics have improved the decoding of motor imagery. Nonetheless, achieving robust, low-latency performance suitable for real-time application in diverse operational environments remains a key challenge. In educational technology, attention-based BCIs have made substantial progress in enabling personalized learning, particularly for students with learning disabilities. However, the scalability of such systems and their integration into formal educational settings still require significant development. In safety-critical industries, such as aviation and maritime operations, EEG-driven attention monitoring has shown promise in managing fatigue and cognitive workload. Despite this potential, ensuring system reliability and generalizability under dynamic, high-stress conditions remains a critical concern.

Building upon these practical experiences and the challenges identified above, future research in attention-based deep learning for BCIs is well-positioned to address current limitations and expand into broader domains with increased precision and robustness. Specifically, attention-enhanced models have the potential to reshape several application areas by leveraging the temporal-spatial sensitivity of EEG signals and the interpretability of attention mechanisms. In clinical neuroscience, attention-guided BCI frameworks can be utilized to develop early diagnostic tools for conditions such as epilepsy, Parkinson’s disease, and Alzheimer's disease. By dynamically attending to frequency bands and electrode regions associated with preictal states, attention-based models can facilitate real-time seizure prediction with fewer false alarms. In assistive technologies, attention mechanisms can enhance brain-controlled exoskeletons and spelling interfaces for individuals with severe motor impairments. Through adaptively focusing on high-informative EEG segments, such as event-related desynchronization (ERD) windows, the system could improve command accuracy under non-stationary conditions. Coupling these models with transfer learning and few-shot adaptation methods may reduce calibration time and personalize device control for locked-in syndrome patients. In mental health monitoring, attention-based emotion recognition systems can be fine-tuned to identify neural signatures of chronic stress, anxiety, or depression. Temporal attention can be applied to isolate emotionally salient EEG segments during resting-state or affective stimulation, facilitating early intervention via wearable neurofeedback devices or closed-loop neuromodulation. In educational neuroscience, future BCI systems can employ attention modules to track cognitive workload in real time, enabling adaptive learning platforms that adjust content difficulty or presentation speed based on attention levels and fatigue. These systems could be particularly beneficial for students with ADHD, where continuous monitoring of frontoparietal EEG activity might allow for real-time feedback or stimulus modulation to improve engagement and learning outcomes. In maritime and aerospace environments, attention-informed BCIs can support neuroergonomic monitoring by detecting shifts in vigilance, workload, or mental fatigue during prolonged or high-stakes operations. By focusing on alpha and theta band dynamics in frontal and parietal regions, these systems can be trained to issue cognitive alerts or task reallocations in response to declining neural performance. Furthermore, embedding these models in low-power wearable headsets with real-time edge inference capabilities is essential for deployment in dynamic, mobility-constrained settings.

Furthermore, future research on attention-based BCI systems should prioritize the development of specialized model architectures that are explicitly aligned with the unique physiological and temporal characteristics of EEG signals. At present, many BCI models are adapted from computer vision and natural language processing, relying on transformations that convert EEG signals into images or sequences. While these approaches have shown promise, they often fail to fully capture the dynamic and non-stationary nature of neural activity. Designing models specifically tailored to EEG data could lead to substantial improvements in clinical neurotechnologies, such as real-time seizure prediction and continuous brain-state monitoring. In addition, it is critical to address the computational inefficiencies and convergence challenges commonly associated with Transformer-based attention mechanisms. These limitations are particularly relevant for mobile neurodiagnostic tools, wearable assistive technologies, and real-time neuroergonomic monitoring systems, where energy efficiency, low latency, and portability are essential. Developing lightweight and optimized attention models will be vital not only for enhancing user experience but also for ensuring reliable real-time inference in resource-constrained environments.

\begin{figure*}
\begin{center}
\includegraphics[width=0.7\textwidth]{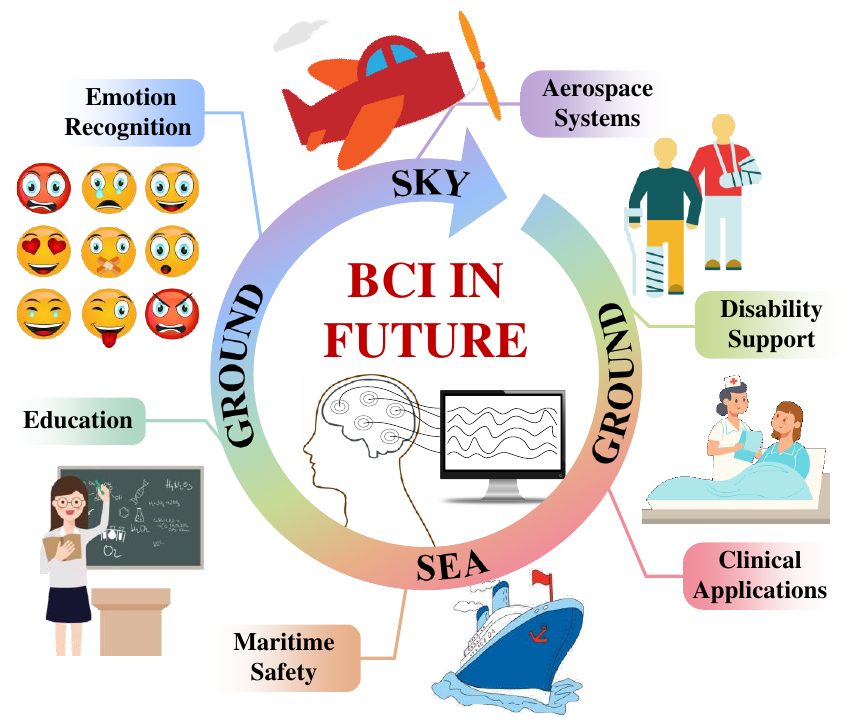}
\end{center}
\caption{Future works of attention modules in BCI.}
\label{fig:future_work}
\end{figure*}

\section{Acknowledgments}
This work was supported by the National Natural Science Foundation of China (62276169), Medical-Engineering Interdisciplinary Research Foundation of Shenzhen University (2023YG004), Shenzhen University Lingnan University Joint Research Programme, Shenzhen-Hong Kong Institute of Brain Science-Shenzhen Fundamental Research Institutions (2023SHIBS0003), the STI 2030-Major Projects (2021ZD0200500), the Open Research Fund of the State Key Laboratory of Brain-Machine Intelligence, Zhejiang University (Grant No. BMI2400008), and the Shenzhen Science and Technology Program (No. JCYJ20241202124222027 and JCYJ20241202124209011).

\bibliographystyle{unsrt}
\bibliography{references.bib}

\end{document}